\documentclass[runningheads,a4paper]{llncs}

\usepackage{amssymb}
\usepackage{graphicx}
\usepackage{subfigure}  % use for side-by-side figures

\usepackage{url}
\urldef{\mailsa}\path|{alfred.hofmann,ursula.barth,ingrid.beyer,natalie.brecht,|
\urldef{\mailsb}\path|christine.guenther,frank.holzwarth,piamaria.karbach,|
\urldef{\mailsc}\path|anna.kramer,erika.siebert-cole,lncs}@springer.com|
\newcommand{\keywords}[1]{\par\addvspace\baselineskip
\noindent\keywordname\enspace\ignorespaces#1}

\newcommand{\picWidth}{0.47}

\begin{document}

\mainmatter  % start of an individual contribution

\title{Session Level Analysis of P2P Television Traces}

\titlerunning{Session Level Analysis of P2P Television Traces}

\author{Arkadiusz Biernacki\inst{1}%
\and Udo R. Krieger\inst{2}}
\authorrunning{Session level analysis of P2P television traces}

\institute{Silesian University of Technology\\
Institute of Computer Science,
44-100 Gliwice, Poland\\
\email{arkadiusz.biernacki@polsl.pl},\\[3mm]
\and
Otto-Friedrich University Bamberg\\
Dep. of Information Systems and Applied Computer Science\\
D-96045 Bamberg, Germany\\
\email{udo.krieger@ieee.org}}


%\toctitle{Lecture Notes in Computer Science}
\tocauthor{Arkadiusz Biernacki, Udo R. Krieger}
\maketitle

\begin{abstract}
In this study we examine statistical properties of traffic generated by the popular P2P IPTV application SopCast. 
The analysis aims at a better understanding of the mechanisms used by such applications and their impact on the network. Since the most popular P2P IPTV applications use proprietary unpublished protocols, we look directly at the generated traffic  focusing on a single session analysis,
 which is the major contribution of our work. We present a basic characterisation of the traffic profile generated by SopCast during every separate session in terms of the intensity, the burstiness, the distribution of the packet sizes and the correlation. We show that some of these statistical properties of the analysed traffic may be quite different depending on the particular session.
\end{abstract}

\section{Introduction}
The use of peer-to-peer overlay networks (P2P) to deliver live television on the Internet (P2P IPTV) is gaining increasing attention. Traditional IPTV service based on a simple unicast approach is restricted to a small group of clients. The overwhelming resource requirement makes this solution impossible when the number of users grows to thousands or millions. By multiplying the servers and creating a content distribution network (CDN), the solution will scale only to a larger audience with regard to the number of deployed servers which may be limited by the infrastructure costs. Finally, the lack of deployment of IP-multicast limits the availability and scope of this approach for a TV service on the Internet scale. Therefore, P2P IPTV has become a promising alternative to IP unicast and multicast delivery protocols.

\begin{figure}%
	\centering
%  	\begin{minipage}{\picWidth\textwidth}%
  		\includegraphics[width=0.7\textwidth]{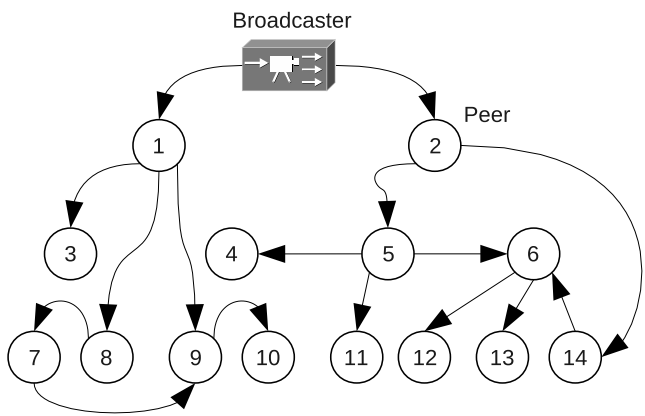}
    		\caption{P2P IPTV mesh architecture.}
		\label{fig:p2pArchitecture}
%		\end{minipage}%
\end{figure}

P2P IPTV (and P2P systems in general) relies on the fact that a set of nodes, called peers, is present at the same time and they act both as clients and servers. Every peer streams media data from multiple neighbouring peers. To coordinate the streaming from multiple sources in a P2P IPTV system, usually a pull-based approach is used. Here a peer collects available data from its neighbours and requests different data blocks from different neighbours. Usually some widely deployed P2P IPTV systems claim to use a mesh-based architecture.
This mesh-based architecture used by P2P IPTV systems is inspired by the BitTorrent system \cite{cohen_bittorrent_2003}. The topology is dynamic and will continuously evolve according to the peering relationship established between peers, see Fig. \ref{fig:p2pArchitecture}.

P2P IPTV traffic can be broadly classified into two categories: signalling and data traffic. The signalling traffic is generated when peer nodes exchange information with each other  regarding the data they possess or any other information required to coordinate the network management. The data traffic comprises mainly audio and video traffic exchanged between peers.
Most of the popular P2P IPTV applications are freely available although their source code is proprietary. In this situation their implementation details and protocols are hidden and unknown to anyone except their developers. Therefore, in order to identify and manage the traffic generated by an P2P IPTV application with proprietary code and protocols, we can only rely on its traffic analysis. The results of this analysis will also provide the necessary data to create appropriate traffic models used in traffic engineering tasks. For our analysis we have chosen SopCast \cite{sopcast_www.sopcast.com_????}. It was developed as student project at Fudan University in China and has become one of the most successful P2P IPTV platforms.

% Update, KR, 14.04.10: splitted
%  	\quad
\begin{figure}%  
\centering
%  	\begin{minipage}{\picWidth\textwidth}%
   		\includegraphics[width=0.7\textwidth]{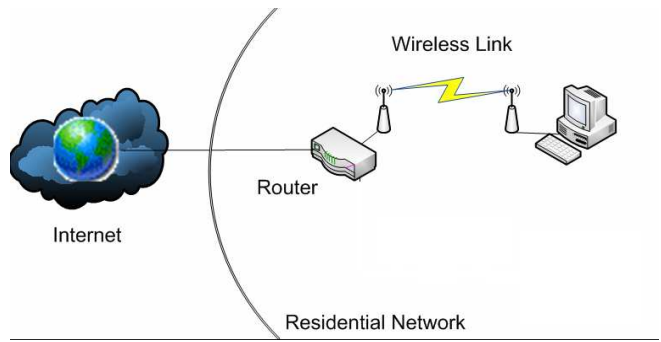}
		\caption{P2P IPTV mesh architecture and test bed of the measurement experiments (see also \cite{Hetnets10}).}%
		\label{fig:labConf}%
%	\end{minipage}%
\end{figure}%

The rest of the paper is organized as follows. After a short discussion of previous work we present our measurement methodology in Section 3.
Then we show the results of our session level analysis of typical  P2P traffic generated by  SopCast in Section 4. Finally, some conclusions are stated.

\section{Previous Work}
In the last years researches have spent a lot of effort to understand   P2P IPTV applications and their protocol internals. However,
to the best of our knowledge, our approach is the first experimental work on P2P IPTV systems exploring properties of the packet traffic (e.g. correlation, packet sizes etc.)  at the session level in greater detail.

The results of several previous studies were obtained by the application of an active crawler to  examine a P2P IPTV system. In this approach the authors try   a partial reverse engineering of the system operations. However, this  methodology is confined by the possibility to break closed and proprietary systems, and we believe that they can be  hardly extended to characterise all the possible P2P IPTV applications. In particular,  \cite{hei_insights_2006} investigates PPLive, whereas \cite{zhang_large_2005} focuses on the commercial re-engineering of Coolstreaming. Both papers mainly provide a big picture of the considered P2P IPTV system, focusing on metrics such as the number of users in the systems, their geographical distribution, the session duration of users, and the distribution of the packet size.

In several other investigations  the authors have focused on the study of specific aspects of a P2P streaming system. For instance, \cite{vu_measurement_2007} gives some preliminary results on the node degrees of popular versus unpopular channels in PPLive. In \cite{wang_stable_2008} the authors analyse the stability of nodes in PPLive and introduce schemes to identify the most stable nodes in the system.

Reverse engineering studies also take into account the quality of service of P2P IPTV systems. The authors of \cite{hei_inferring_2007} explore how to monitor remotely the network-wide quality in PPLive systems exploiting buffer maps. The latter summarise the chunks of data that the systems  have currently cached and made available for sharing.
The authors of \cite{agarwal_performance_2008} yield several quantitative measures of P2P video streaming multicast sessions exploiting logs which were provided by an unspecified commercial P2P streaming system.

In \cite{ali_measurement_2006} the authors analyse and compare PPLive and SOPCast investigating the control traffic, resource usage and locality as well as the stability of the data distribution.   \cite{silverston_traffic_2009} presents a comparative evaluation of four commercial systems, namely PPLive, PPStream, SOPCast and TVAnts, by measuring and analysing their network traffic.
These mentioned systems, except PPStream, are analysed in \cite{Horvath_dissecting_2009}, too. The study has investigated the properties of the algorithms driving the P2P data exchange including the bias with respect to the peer bandwidth, the exploitation of peer locality information, and the presence of incentive mechanisms that govern the data exchange. At last,  an experimental analysis of PPLive and Joost is presented in \cite{ciullo_understanding_2008} to examine the characteristics of both data and signalling traffic distribution, while the authors of \cite{alessandria_p2p-tv_2009}  reveal how P2P IPTV applications cope with changing network conditions (delay, loss and available capacity) by observing the received bitrate and the number of contacted peers.

\section{Methodology and Measurements of Live-Streamed SopCast Traffic}
To outline precisely the scope of the measurement and the analysis of traffic, we have defined several metrics:
\begin{itemize}
\item
As \textbf{session traffic} we have defined a flow, i.e.   an exchange of TCP or UDP packets between source host A and destination host B which is represented by a four-tuple $\{IP_A, P_A, IP_B, P_B\}$. Here $IP_X$ and $P_X$ denote the IP address and port number of host
$X \in \{A, B\}$, respectively. Each flow has a direction, thus the two flows $\{IP_A, P_A, IP_B, P_b\}$ and $\{ IP_B, P_B, IP_A, P_A\}$ are treated as two separate streams.

\item
The
\textbf{session duration} is defined as the time between the first packet and the last packet of a session. Note that this definition includes cases where a flow stops withing the session for a period of time and starts again. We call the latter packet streams  micro-flows.

\item
The
\textbf{traffic intensity} is defined here as the number of packets which arrive in a particular unit of time.

\item
The
\textbf{ Index of Dispersion for Counts} (IDC) denotes the variance of the number of arrivals in a time unit divided by the mean number of arrivals in that time unit. IDC characterises the burstiness of the traffic intensity. It is relatively straightforward to estimate it, cf. \cite{gusella_characterizingvariability_1991}.
\end{itemize}

Since the SopCast system uses proprietary protocols and very little is known about their structure, we have performed a passive measurement study at the Computer Networks Laboratory of Otto-Friedrich University Bamberg, Germany, during the second quarter of 2009 and analysed the
collected  cleaned  traces of representative SopCast sessions.
Carrying out the study in a typical home environment with an ADSL access to the Internet (see Fig. \ref{fig:labConf},
cf. \cite{Hetnets10}), we have tuned on the client to a unique Chinese sport channel and recorded live streams of a soccer match during half an hour.
According to our previous insights \cite{Hetnets10}, the latter data set is able to reflect the major teletraffic features of the basic P2P mechanisms of SopCast sessions.

The test bed provides an asymmetric Internet access of a representative residential network with a maximal download rate of 6016 kbps and an upload rate of 576 kbps. However, according to our experience the actual ADSL rates are smaller due to the impact of damping on the access line and multiplexing effects.
In the deployed wireless application scenario the monitored client is running on a desktop PC IBM Thinkcentre with 2.8 GHz Intel Pentium 4 processor, 512 MB RAM, and OS Windows XP Home. It has been attached to the corresponding ADSL router by a Netgear WG111 NIC operating the IEEE802.11g MAC protocol over a wireless link.
The measurement sessions have typically generated traces with a volume of 140 MB and around 373 thousand IP packets.

\section{Measurement Results}
\begin{figure}%
	\centering
%  	\begin{minipage}{\picWidth\textwidth}%
  		\includegraphics[width=0.9\textwidth]{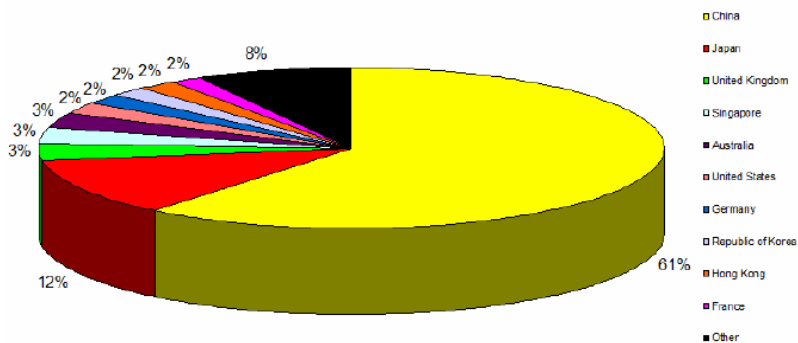}
    		\caption{Geographical peer distribution of a typical WLAN session.}
		\label{fig:Geo-df-1}
%		\end{minipage}%
\end{figure}

The preprocessing of the traces has revealed that SopCast relies on UDP packet transfers to realise its
major signalling and content transport functions. During such a typical SopCast session only 0.05\% TCP packets
have been recorded. The upload/download ratio has been about 1:4 and more than 1000 ports have been utilised to transfer the UDP flows within the established overlay network of feeding peers during a session (cf. \cite{Hetnets10}).
 The majority of these feeders, typically more than 60\%, stems from China as expected, see Fig. \ref{fig:Geo-df-1}.

\begin{figure}%
%  	\qquad
  \centering
%  	\begin{minipage}{\picWidth\textwidth}%
   		\includegraphics[width=0.9\textwidth]{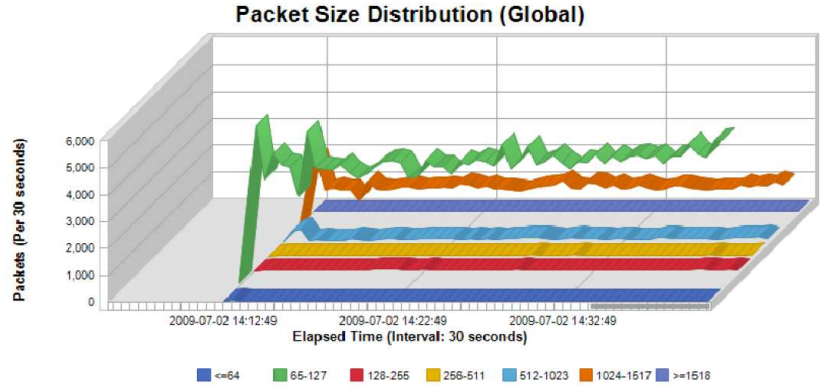}
		\caption{Frame size distribution realized by a typical SopCast session.}%
		\label{fig:frame-size-1}%
%	\end{minipage}%
\end{figure}

Plotting the frame length distribution over time reveals the request-response pattern of the pull-mesh approach applied by SopCast, see Fig. \ref{fig:frame-size-1}. Signalling packets mainly generate frame lengths in the range of 67 to 127 bytes whereas content traffic is reflected by frame lengths in the interval $[1024, 1500]$ bytes. A deep packet inspection shows that the video chunks are normally represented by a UDP payload of 1320 bytes, i.e. 1362 byte long frames.

% Update KR, 14.04.10: splitted

%------------------------------------------------------------------

\begin{figure}[h]
\begin{center}
\subfigure[Downloaded traffic.]{\label{fig:sessionDurDown} \includegraphics[width=\picWidth\textwidth]{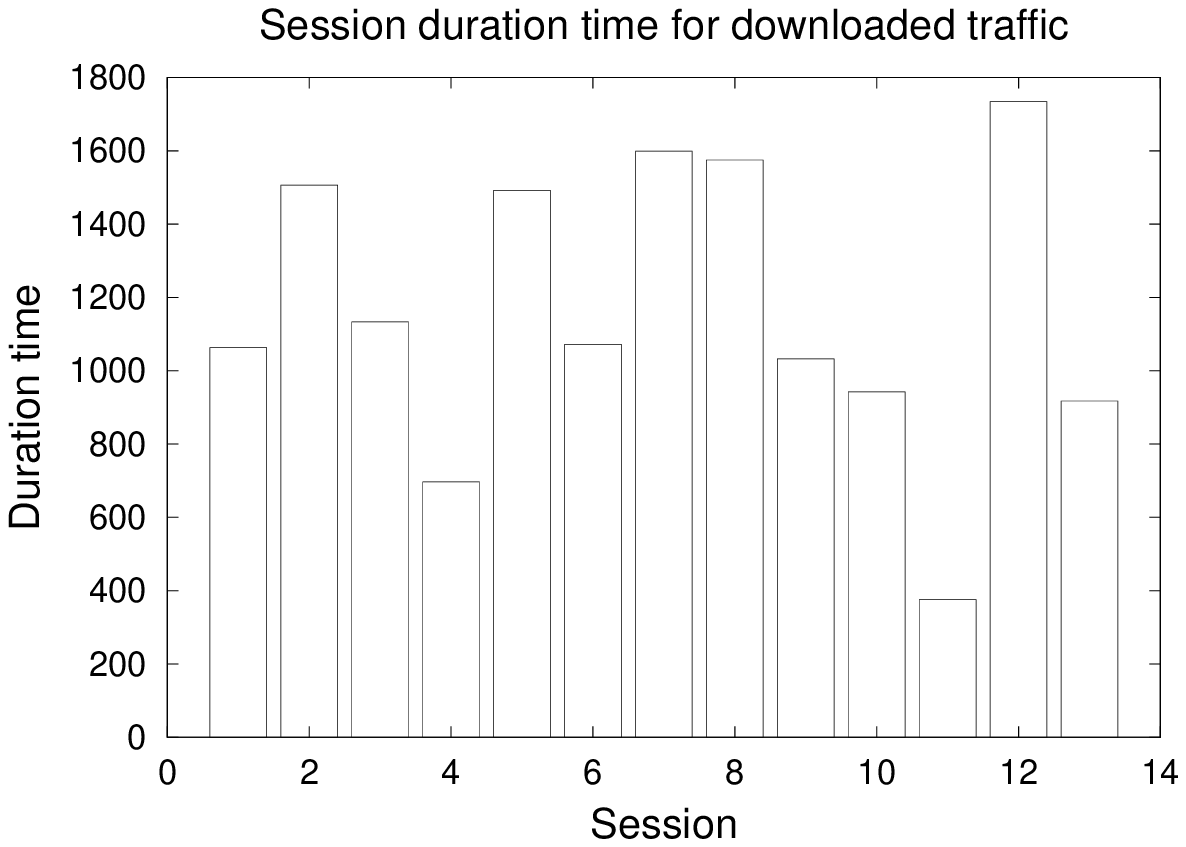}}
\quad
\subfigure[Uploaded traffic.]{\label{fig:sessionDurUp} \includegraphics[width=\picWidth\textwidth]{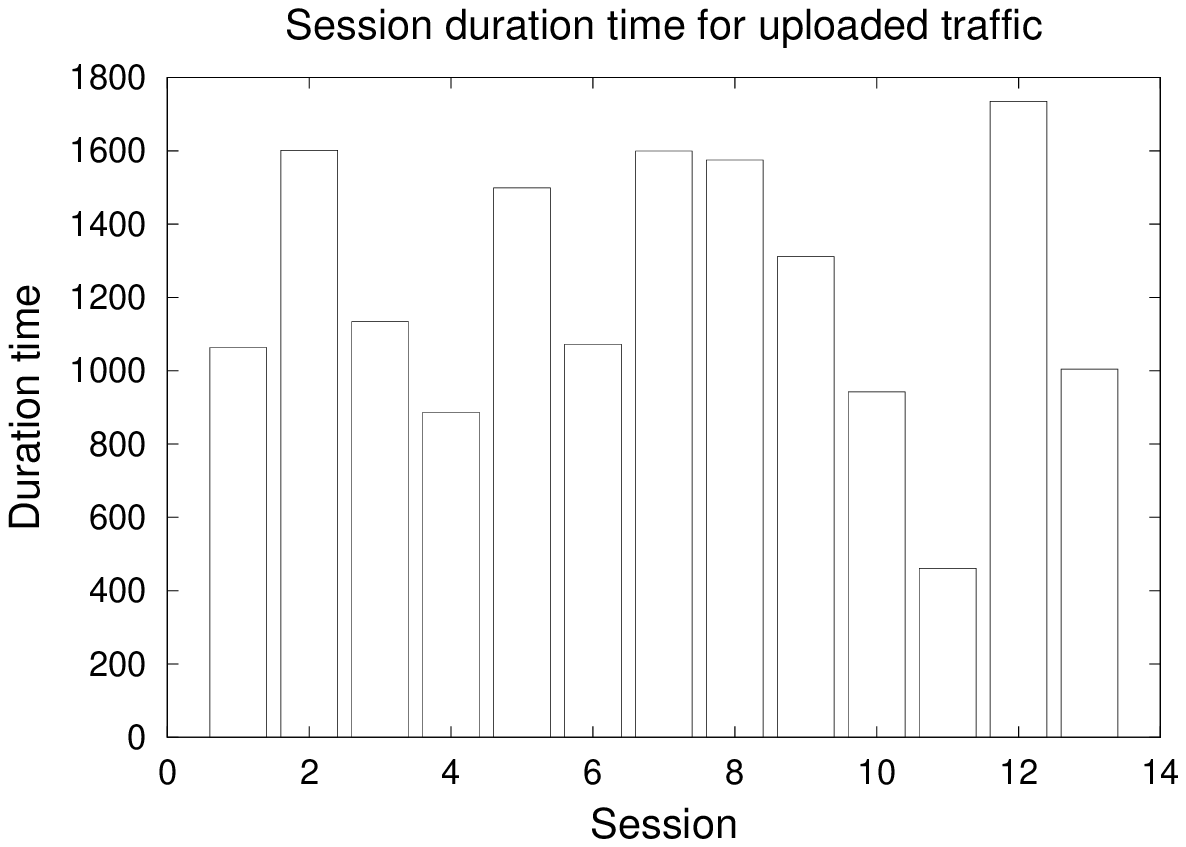}}
\caption{Session durations [s] of SopCast traffic.}
\label{fig:sessionDur}
\end{center}
\end{figure}

\begin{figure}
\begin{center}
\subfigure[Downloaded traffic.]{\label{fig:downTrafficPlot} \includegraphics[width=\picWidth\textwidth]{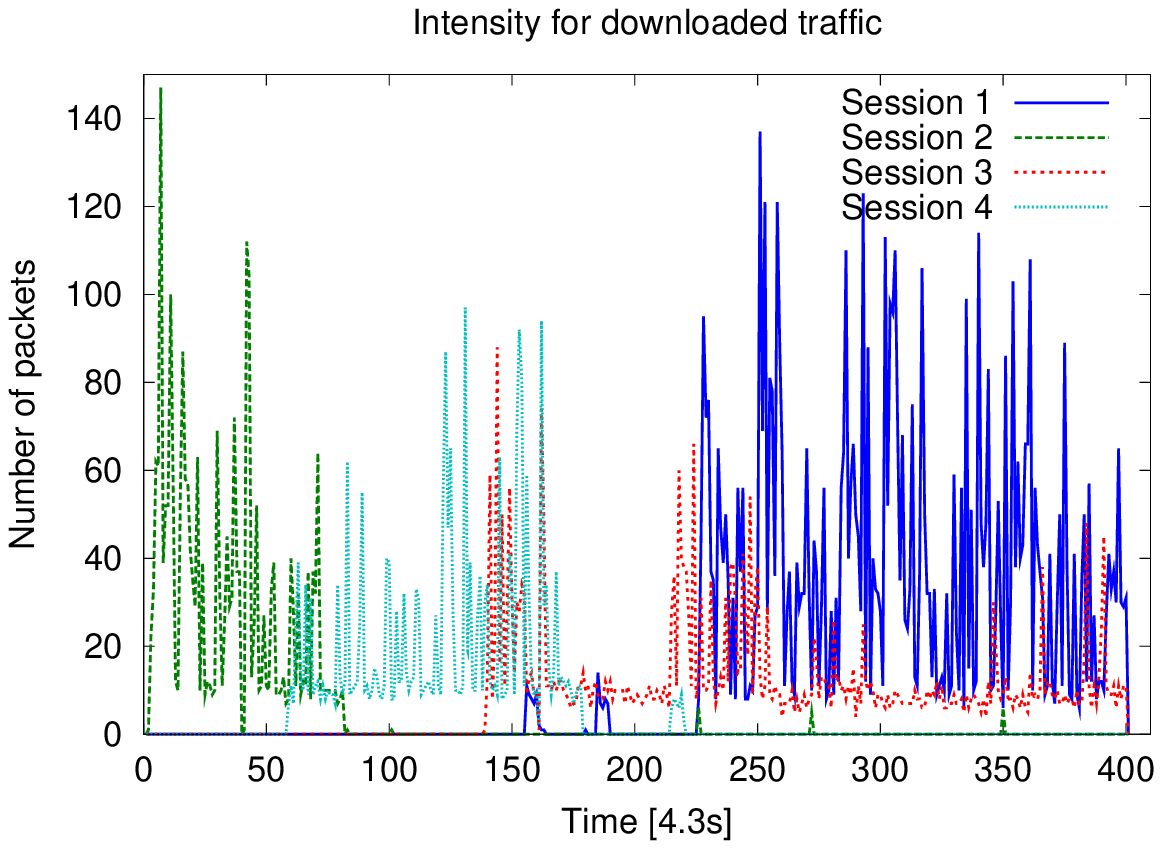}}
\quad
\subfigure[Uploaded traffic.]{\label{fig:upTrafficPlot} \includegraphics[width=\picWidth\textwidth]{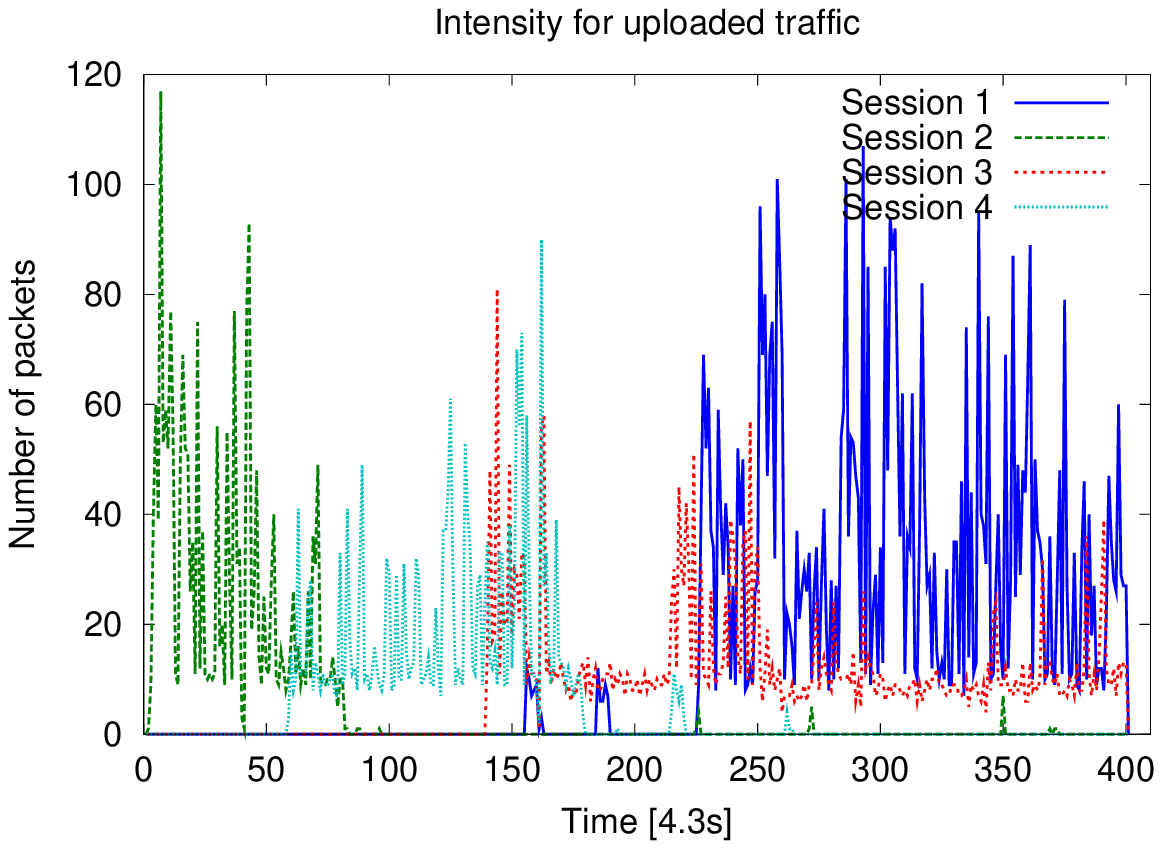}}
\caption{Traffic intensity in time domain.}
\label{fig:TrafficPlot}
\end{center}
\end{figure}

From the set of several dozen session traces 
% for the further analysis 
we have chosen several ones taking into account their length which last for   most of the sessions between 15 [min] and 30 [min], see Fig. \ref{fig:sessionDur}. Since we did not separate signalling traffic from data traffic, the communication is bidirectional. Thus, the durations of the  sessions in the upload and download directions are highly correlated and both  figures are nearly similar.
Analysing the session flows, the resource usage in terms of network bandwidth is an important metric. The traffic intensity in the time domain is presented in Fig. \ref{fig:TrafficPlot}. On the basis of the visual assessment of the session traces we may claim that their nature is quite bursty.

\begin{figure}
\begin{center}
\subfigure[Uploaded traffic.]{\label{fig:meanUp} \includegraphics[width=\picWidth\textwidth]{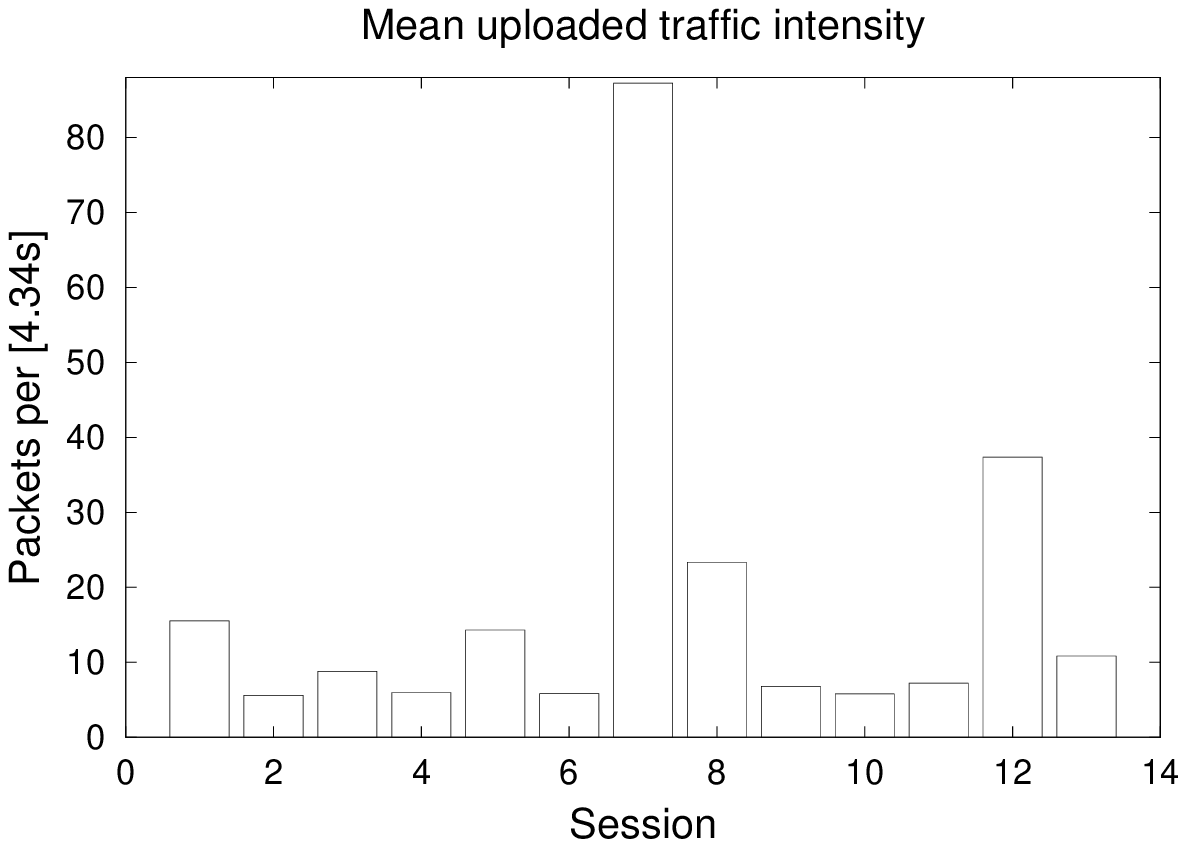}}
\quad
\subfigure[Downloaded traffic.]{\label{fig:meanDown} \includegraphics[width=\picWidth\textwidth]{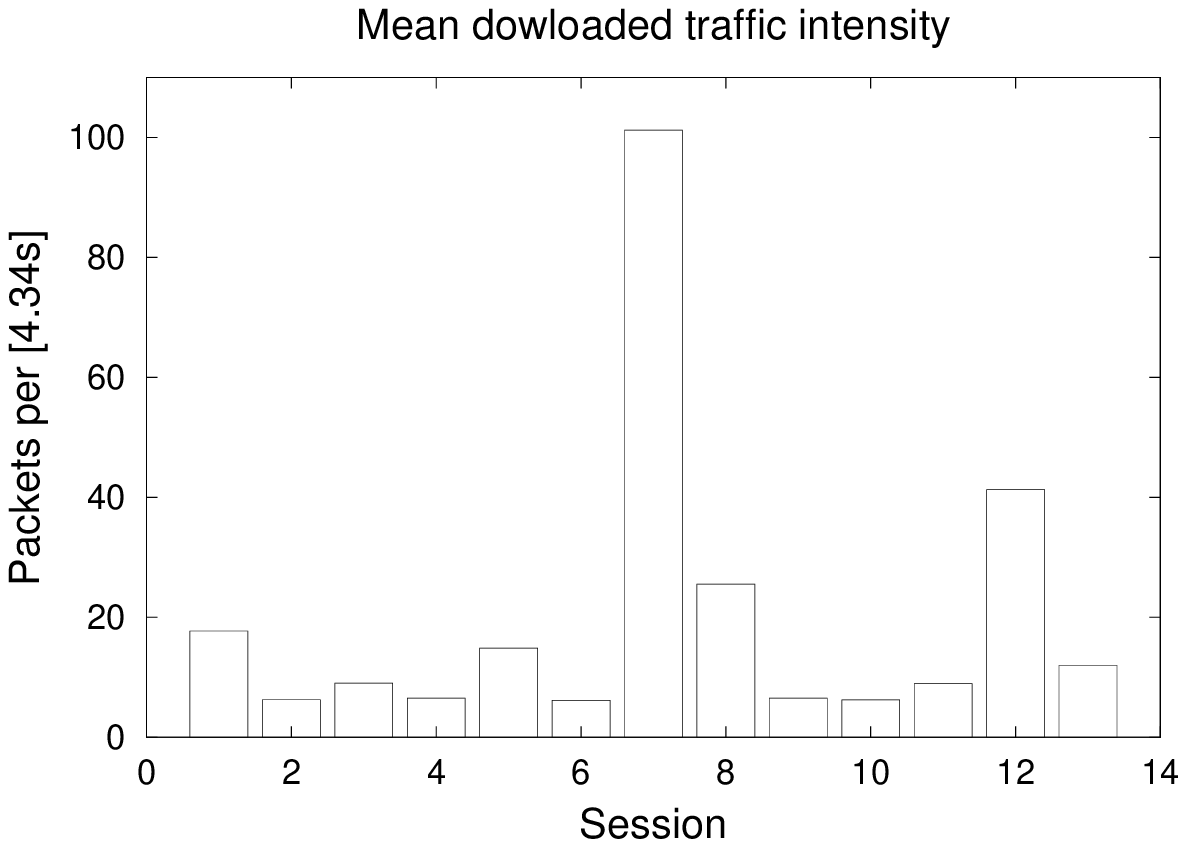}}
\caption{Mean traffic intensity.}
\label{fig:meanTraffic}
\end{center}
\end{figure}

The mean traffic intensity is presented in Fig. \ref{fig:meanTraffic}. The intensity is quite different for each session and, as one can see, one of the neighbouring peers is particularly active in terms of the amount of exchanged data.

% Correction, KR, 14.04.10

\begin{figure}
\begin{center}
\subfigure[Downloaded traffic.]{\label{fig:dispDown} \includegraphics[width=\picWidth\textwidth]{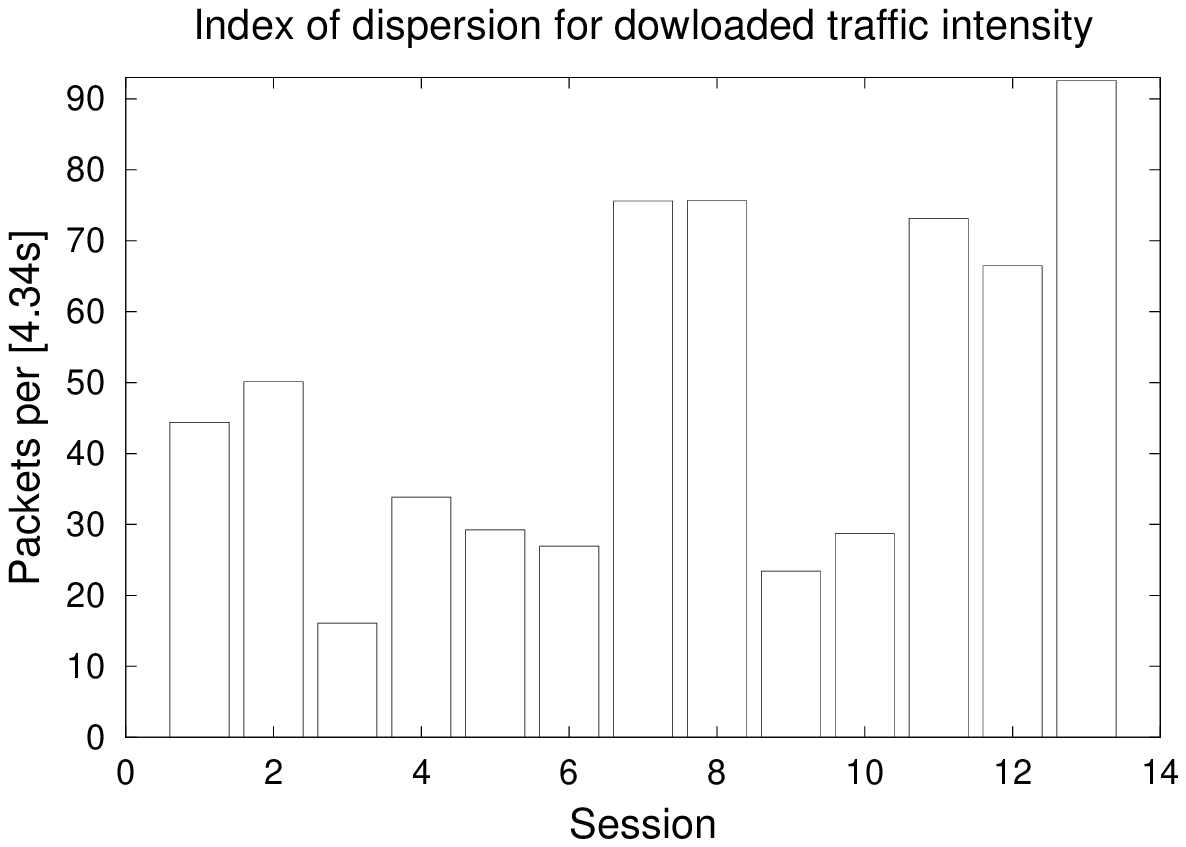}}
\quad
\subfigure[Uploaded traffic.]{\label{fig:dispUp} \includegraphics[width=\picWidth\textwidth]{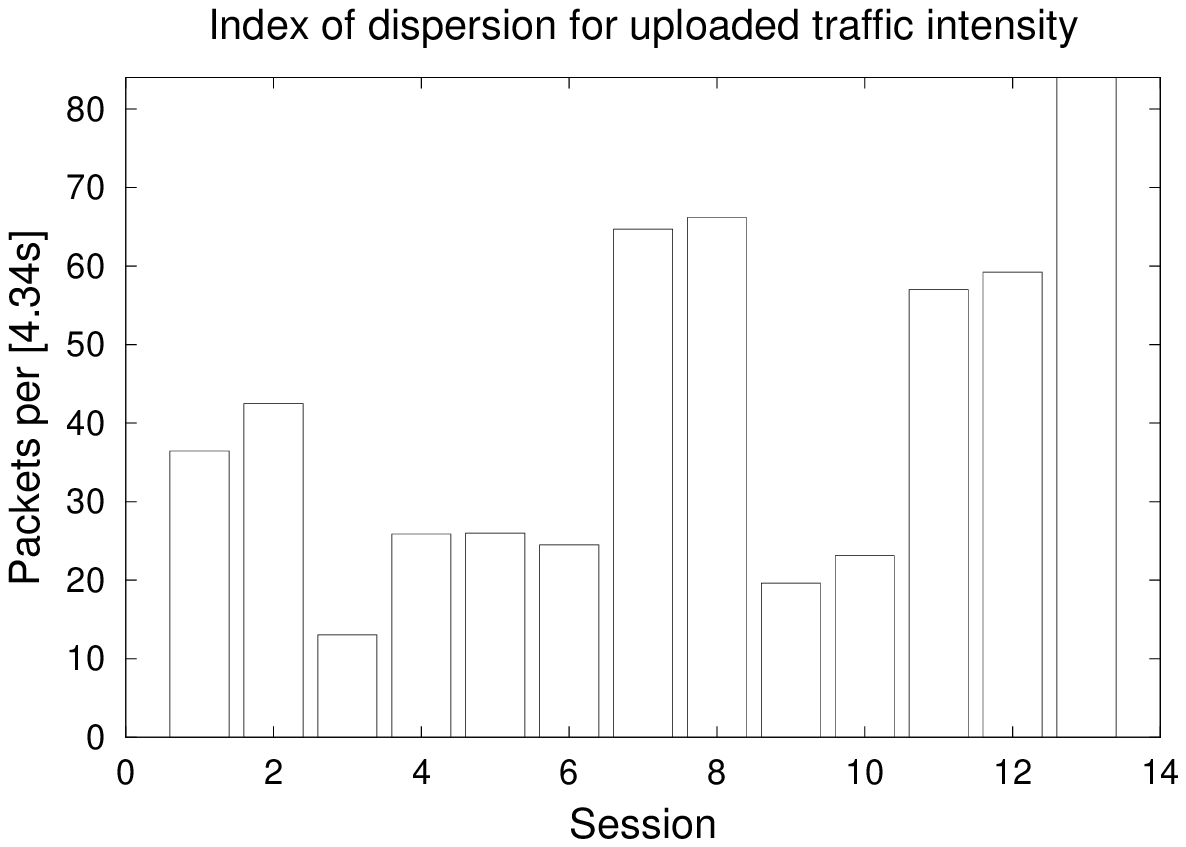}}
\caption{IDC of the traffic intensity.}
\label{fig:dispTraffic}
\end{center}
\end{figure}

Relying on the plot of the IDC, cf. Fig. \ref{fig:dispTraffic}, we may state that the burstiness of the traffic varies heavily depending on the session and its spread is nearly of the order of one magnitude.
 On a first glance, the IDC does not  seem to be correlated, neither with the traffic intensity (see Fig. \ref{fig:meanTraffic}) nor its duration time (see Fig. \ref{fig:sessionDur}).

\begin{figure}
\begin{center}
\subfigure[Downloaded traffic.]{\label{fig:down3dhist} \includegraphics[width=\picWidth\textwidth]{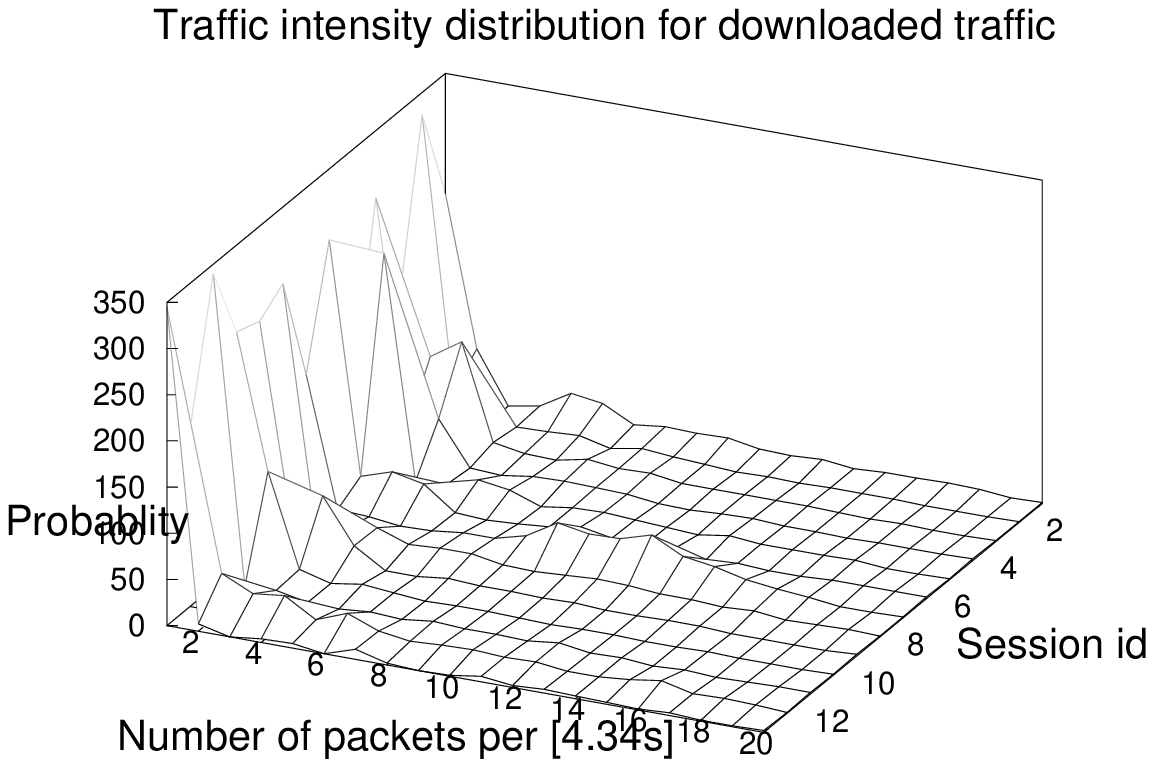}}
\quad
\subfigure[Uploaded traffic.]{\label{fig:up3dhist} \includegraphics[width=\picWidth\textwidth]{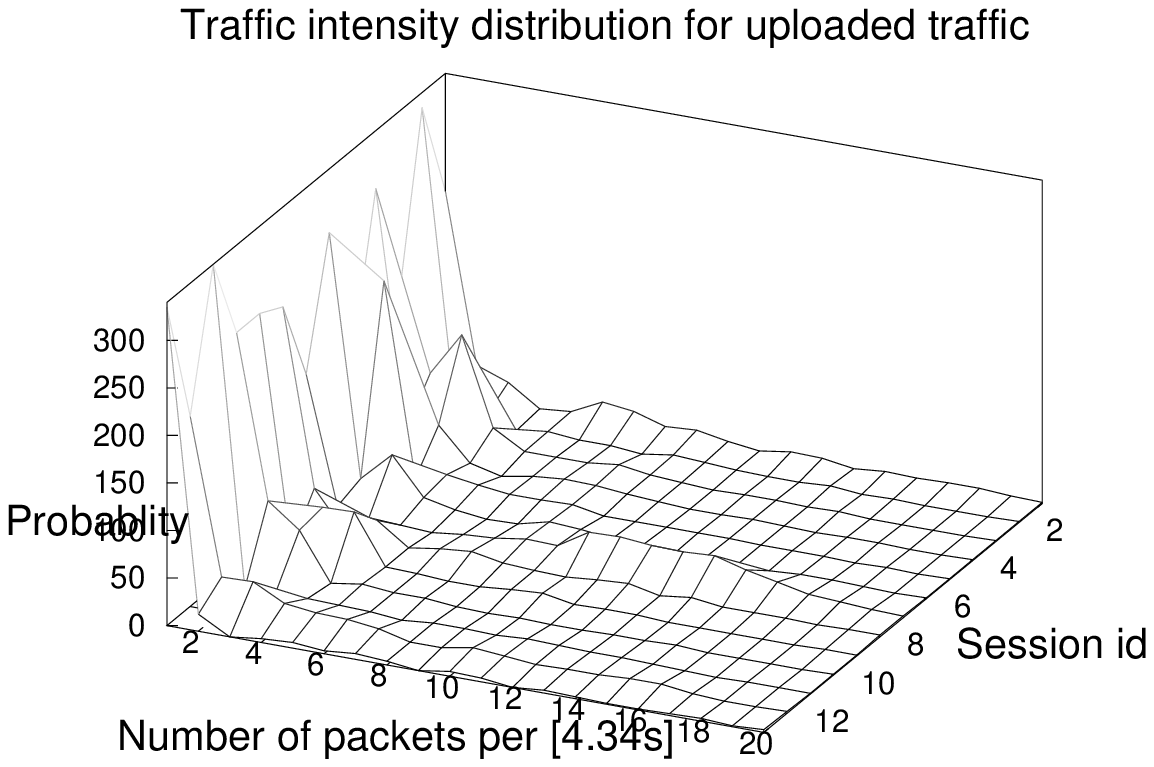}}
\caption{Distribution of the traffic intensity.}
\label{fig:trafficHist}
\end{center}
\end{figure}

The distribution of the traffic intensity is presented in Fig. \ref{fig:trafficHist}. Due to the bursty nature of the traffic (compare Fig. \ref{fig:TrafficPlot}) there are many epochs where no traffic is transmitted, thereby creating an ON/OFF structure in the traffic pattern. Interestingly the number of the OFF periods is roughly similar for all examined session flows. Furthermore, for some sessions we can observe so called long tails in the distribution. It is especially visible in the case of the flow with the highest traffic intensity (compare with  Fig. \ref{fig:meanTraffic}). Although the downloaded traffic \ref{fig:down3dhist} comprises   signalling and audio-visual traffic  in contrast to the uploaded traffic \ref{fig:up3dhist}, which comprises mainly signalling traffic, the distributions of both types are visually similar.

\begin{figure}[h]
\begin{center}
\subfigure[Downloaded traffic.]{\label{fig:downPak3dhist} \includegraphics[width=\picWidth\textwidth]{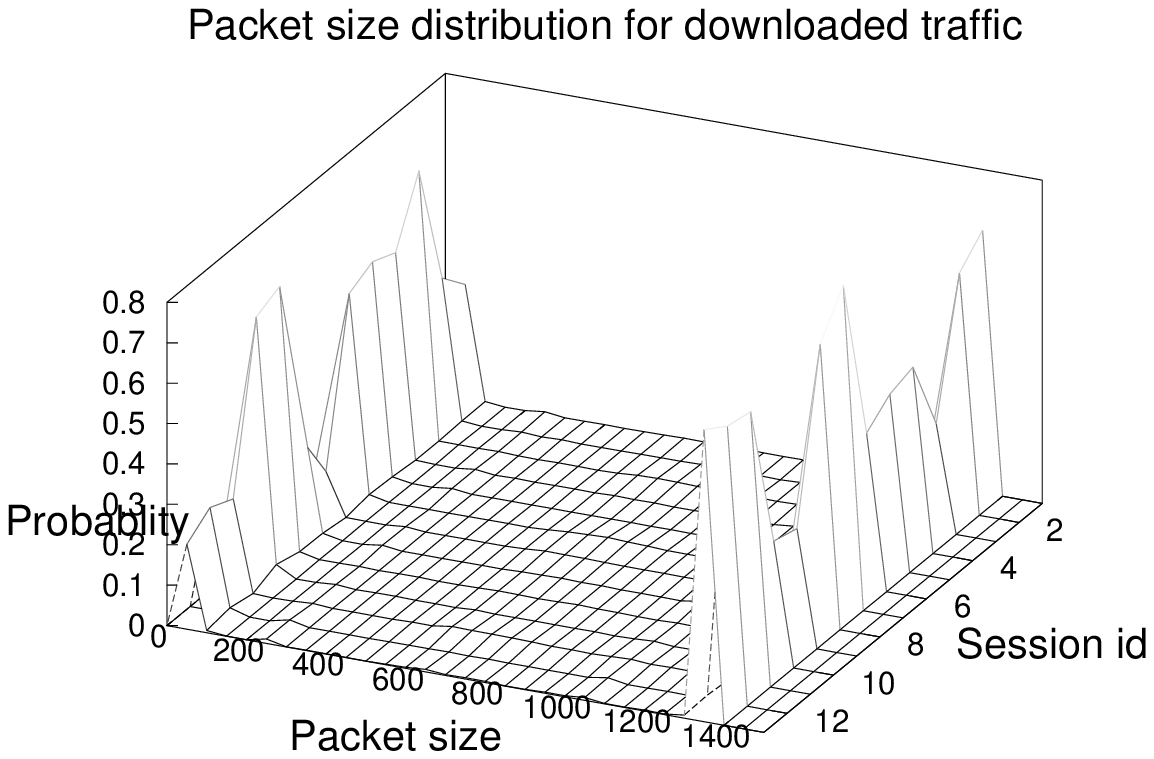}}
\quad
\subfigure[Uploaded traffic.]{\label{fig:upPak3dhist} \includegraphics[width=\picWidth\textwidth]{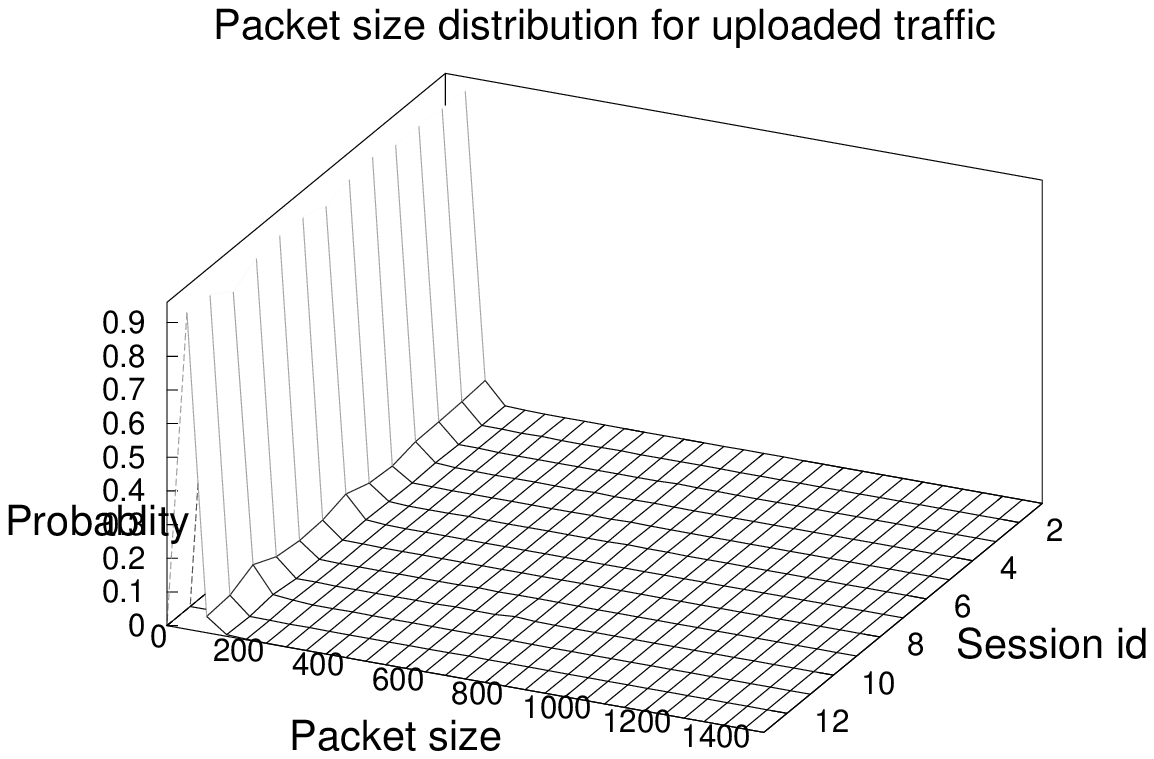}}
\caption{Multi-session packet size distribution.}
\end{center}
\end{figure}

 In case of the downloaded traffic the packet sizes within each session, depicted in Fig. \ref{fig:downPak3dhist}, may be described by bimodal distribution with local peaks around 100 bytes and 1300 bytes, respectively, for signalling and audio-video traffic. These findings are in agreement with previous studies, cf.   \cite{ciullo_understanding_2008}, \cite{silverston_traffic_2009}. However, it can be seen that 
 in the terms of the number of packets the share of signalling traffic  to audio-video traffic may heavily vary depending on the session.

In contrary, in case of the uploaded traffic the packet sizes are distributed unimodally with peeks around 100 bytes, thereby suggesting a unidirectional flow of our audio-video traffic.

\begin{figure}[h]
\begin{center}
\subfigure[Download]{\label{fig:downCorr} \includegraphics[width=\picWidth\textwidth]{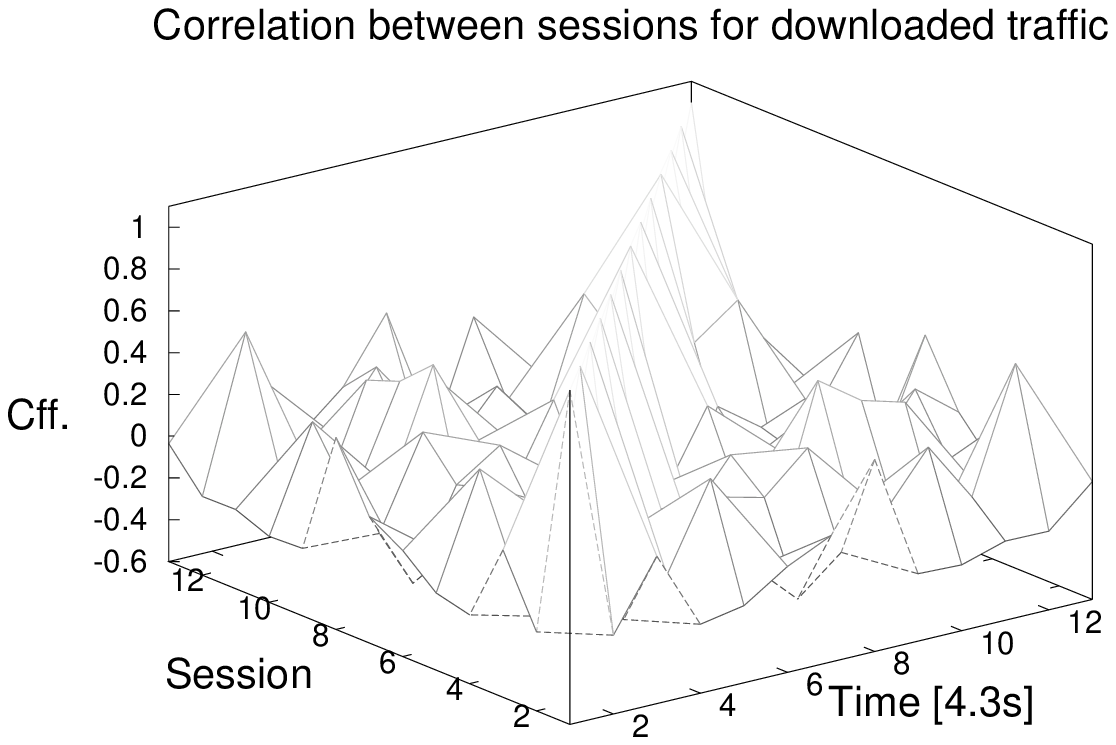}}
\quad
\subfigure[Upload]{\label{fig:upCorr} \includegraphics[width=\picWidth\textwidth]{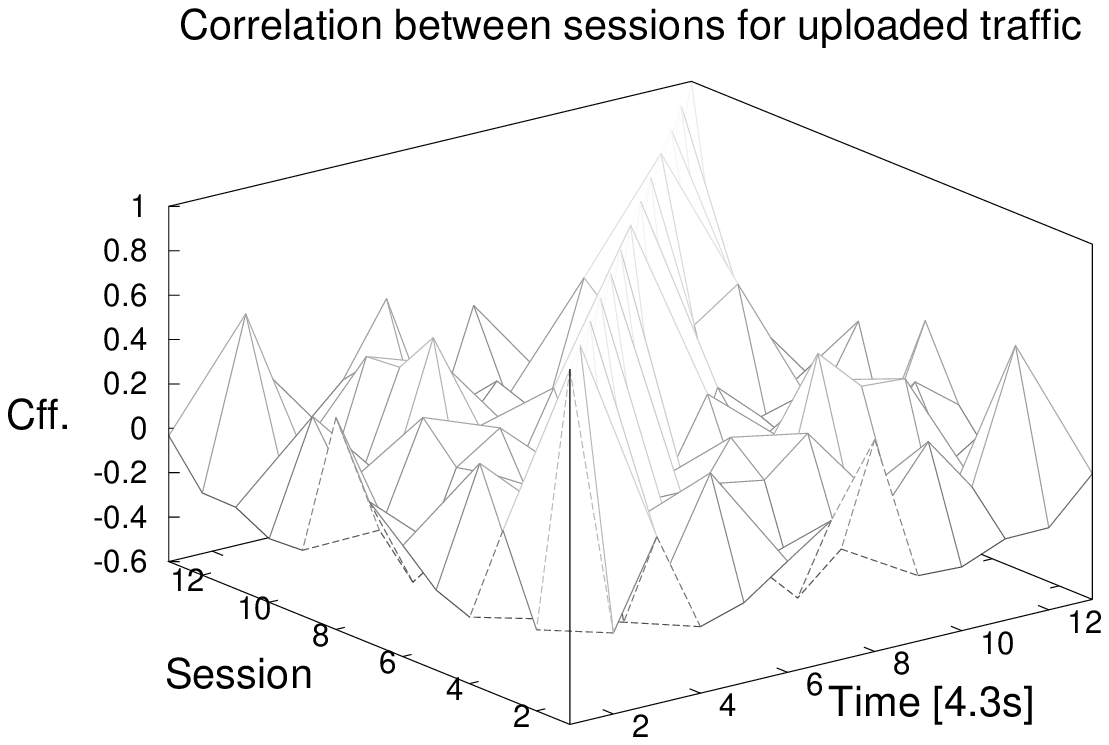}}
\caption{Correlation between multi-session traffic.}
\label{fig:Corr}
\end{center}
\end{figure}

We can further observe    both for the downloaded and the uploaded traffic
 that there is a moderate positive or negative correlation between the traffic intensity generated by some of the individual sessions, see Fig. \ref{fig:Corr}. It is worth noting that during the collection of the traces  the available network bandwidth for the P2P IPTV application remained constant. Thus, the   correlation coefficient between $-0.4$ and $+0.4$ could  perhaps be explained by churn of the neighbouring peers.

\section{Conclusions}
In the paper we have presented an overview of the popular P2P IPTV technology. By means of  passive measurements we have characterised P2P IPTV traffic generated by one of the most popular P2P IPTV  systems, namely SopCast. We have presented a basic characterisation of the traffic profile generated during every session in terms of the intensity, the burstiness, the distribution of the packet sizes and the correlation. 

On the basis of our analysis we may state that the traffic generated by SopCast is quite bursty and that the traffic intensity in some of the analysed sessions may be described as heavy tailed. The correlation of the traffic intensity  between particular sessions is moderate without negative or positive bias.  For every session the packet sizes are bimodally distributed with local peaks for signalling and audio-video traffic.

Our measurement results intend to provide a better understanding of the best way to design a successful large scale P2P IPTV system. Insights obtained in this study should be valuable for the development and deployment of future P2P IPTV systems which are performed by related EU projects like NAPA-Wine and P2P-Next.

To obtain a stronger in-depth look onto the teletraffic features of   P2P IPTV traffic,  we plan in the near future to separate signalling and data traffic and to analyse the statistical properties of the  micro-flows within a session in a thorough statistical way.
Preliminary results on the statistical traffic characterisation have already been provided by
\cite{Hetnets10} and \cite{Markovich-WWIC10}.

\subsection*{Acknowledgement}
The authors acknowledge the   support by   the projects BMBF MDA08/015 and COST IC0703.
They also express their sincere thanks to P. Eittenberger who has collected  the SopCast traces.

% Update KR, 14.04.10
%\bibliographystyle{plain}
%\bibliographystyle{splncs}
%\bibliography{bib}

\end{document}